# Gas Eruption Phenomenon Happening from Ga-In Alloy in Electrolyte

Ruiqi Zhao,[1,2] Hongzhang Wang,[1,3] Jianbo Tang,[1,3] Wei Rao,[1,4,a)] and Jing Liu,[1,3,4,a)]

[1] *Beijing Key Lab of CryoBiomedical Engineering and Key Lab of Cryogenics, Technical Institute of Physics and Chemistry, Chinese Academy of Sciences, Beijing 100190, China*

[2] *Department of Chemistry, University of Chinese Academy of Sciences, Beijing 100049, China*

[3] *Department of Biomedical Engineering, School of Medicine, Tsinghua University, Beijing 100084, China*

[4] *College of Future Technology, University of Chinese Academy of Sciences, Beijing 100049, China*

[a)] Authors to whom correspondence should be addressed. E-mail: weirao@mail.ipc.ac.cn; jliu@mail.ipc.ac.cn.

**Abstract:** We report a gas eruption phenomenon caused by electrolysis of liquid Ga-In alloy in an electrolyte, especially NaOH solution. A volcanic eruption-like blowout of gas occurred from the orifice on the alloy surface. In addition to gas plume, large gas bubbles were also generated and the total gas yield increased as In ratio was increased. It is found that destructiveness of the passivation layer on the Ga-In alloy is critical to gas generation. The mechanism of gas eruption can be ascribed to a galvanic interaction happens owing to passivation film and alloy with different activity connected as electrode in electrolyte. Further investigation demonstrated that the lattice of the film expands because of the incorporation of indium, which brings about the decrease in band gap and finally enhances more gas generation. These findings regain the basic understanding of room temperature liquid metal inside electrolyte.

Recently, studies on gallium-based liquid metals have drawn considerable attention, particularly for their potential applications in hydrogen generation. Such alloys have in fact been the focus of many studies. The addition of aluminum to Ga alloy contributes to self-driven motions.[1-7] The majority of studies have proven the addition of aluminum or other metal particles as the main regulator of hydrogen generation.[3, 8-14] The generation of gas is considered to be a propelling force of a bunch of phenomena, including the motion of Ga-In-Al in aqueous NaOH solutions,[15] and the oscillation phenomenon of a copper wire embedded inside Ga-In-Al self-powering system.[16] Alternatively, Ga–In liquid metal can be applied versatile to self-healing or contrast enhancement.[17] Tang et al. showed that hydrogen could be generated from a drop of Ga-In alloy when left in contact with solid metal particles.[18] However, phenomenon of gas evolution from bulk Ga-In alloy has always been overlooked and no studies have yet been reported.

In this letter, we report an interesting gas eruption phenomenon occurring in Ga-In alloy when it contacts with an electrolyte. Over the experiment, 10 cm petri dish was pre-filled with 50 ml Ga-In alloy in the open air [Fig.1(a)], then 30 ml NaOH electrolyte was transferred onto the surface of alloy. When immersed in a 1 M alkaline electrolyte, a number of gas plumes appeared at the surface of GaIn$_{10}$ (Ga : In = 9:1,w/w, see Fig.1(b) & Multimedia view), GaIn$_{24.5}$ (Ga : In = 75.5:24.5, w/w, see Fig.S1& supplemental material S1) and GaIn$_{50}$ alloy (Ga : In =1:1, w/w, see Fig.1(c) & Multimedia view). The eruption phenomena have been recorded with a Canon EOS 70D camera with macro lenses. The experiments were performed at a temperature of 22°C.

The electrolyte activates the surface layer of Ga-In alloy and creates many orifices [Fig.1(d), side view in supplemental material S2], from which gas column pours out. In addition, the hydrogen gas generation is rapid and constant, and the continuous gas flow forms a gas plume in the electrolyte (see Multimedia view). This is a similar phenomenon to that of volatile gas blended with magma erupting from the seabed, which then rises to the top of sea water. Considering the similarities with submarine volcanic eruptions, we divide our eruption column into three zones[19], namely regions of jet, ascent, and diffusion, respectively [Fig.1(e)]. The jet region is located at the lowest part of the eruption column, and the column width is approximately 50~200 μm. The gas flow from the orifice is sufficiently powerful to overcome the frictional resistance, and the velocity at the alloy/electrolyte interface is controlled mainly by the volatile gas release of the Ga-In alloy. In the ascent region (upper zone), the speed of the gas column is slower than that in the jet phase. The gas in the column scatters outward, and the force driving the upward motion is dominated by buoyancy. The diffusion region is located at the top of the eruption column where the eruption column rises and spreads in the horizontal direction. In diffusion region, the pressure of the eruption column and the external atmosphere reaches equilibrium.

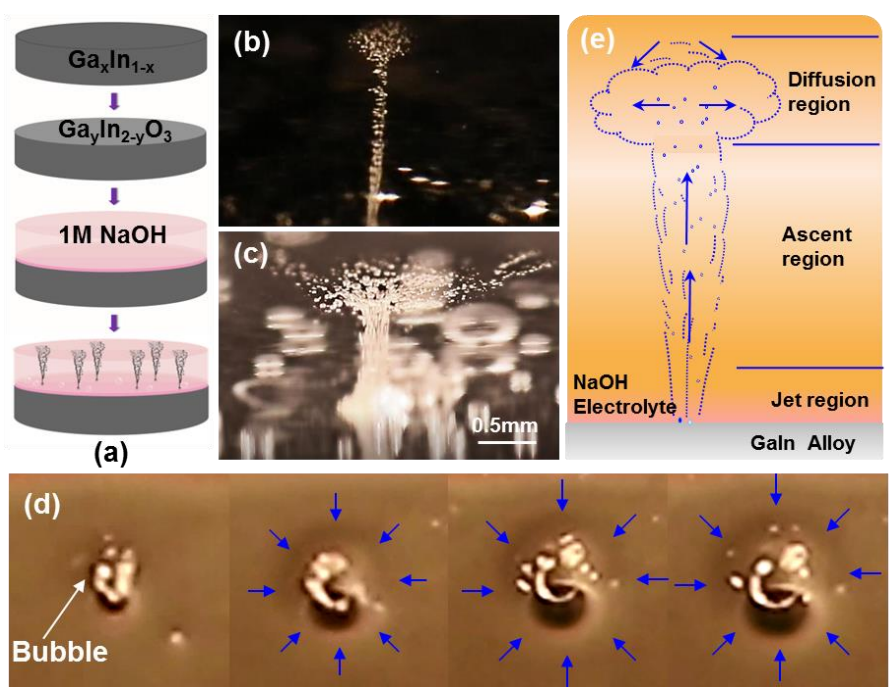

FIG.1. Images of gas plumes and bubbles appearing on the interface of $GaIn_{10}$ alloy and $GaIn_{50}$ alloy in 1 M NaOH solution (side view). (a) Experimental diagram of the formation of gas plumes and bubbles. x: 1~100%; y: 1 or 2. (b) Enlarged image of a typical gas plume (side view) showing gas dispersion from $GaIn_{10}$. A reflection of the plume also appears on the surface of the liquid metal. (Multimedia view) (c) Enlarged image of typical gas plume and big bubbles (side view) rising through the surface of $GaIn_{50}$ alloy. (Multimedia view) (d) A gas bubble emerges from an orifice in GaIn24.5 (front view), the region of orifice is circled by blue arrow. White arrow: gas bubble, Blue arrow: orifice, Background: $GaIn_{24.5}$.(e) Schematic

illustration of a plume rising through the surface of GaIn$_{10}$ alloy, and three parts of the eruption column.

Expanded images show not only gas plumes, but also gas bubbles generate, as the ratio of Indium (In) is increased in the alloy. To allow a comparison, three types of Ga-In alloy, GaIn$_{10}$, GaIn$_{24.5}$, GaIn$_{50}$ are used in this study. Their reaction with 1 M NaOH electrolyte is fully shown in Fig. 2. The volcanic eruption phenomenon of GaIn$_{10}$ in 1 M NaOH solution is illustrated in Fig. 1(b) and Fig. 2(a). Gas plumes and rare bubbles are released through several orifices that appear at the interface of the liquid metal and electrolyte. As the ratio of In increases, the speed of the gas blown out from GaIn$_{24.5}$ in 1 M NaOH solution increases and more gas plumes form, as shown in Fig. 2(b). Furthermore, gas bubbles are released directly to the upper surface of the NaOH solution, driven by the gas flow (Fig. S1). The bubbles fracture into several smaller bubbles at the surface of the electrolyte. Fig.2(c) demonstrates that the gas generation phenomenon of the GaIn$_{50}$ in contact with the 1 M NaOH solution is more vigorous compared with that from GaIn$_{10}$ and GaIn$_{24.5}$. For GaIn$_{50}$ in 1 M NaOH solution, large bubbles remained on the surface of the liquid metal. We found that increasing the proportion of In in the Ga-In alloy led to a faster gas plume and more bubbles [illustrated in Fig. 2(d)~(f)]. The quantitative gas evolution rate is presented in Fig. 2(g). It is clear that GaIn$_{50}$ displays highest gas production performance, which is consistent with our observations of the gas plumes and bubbles.

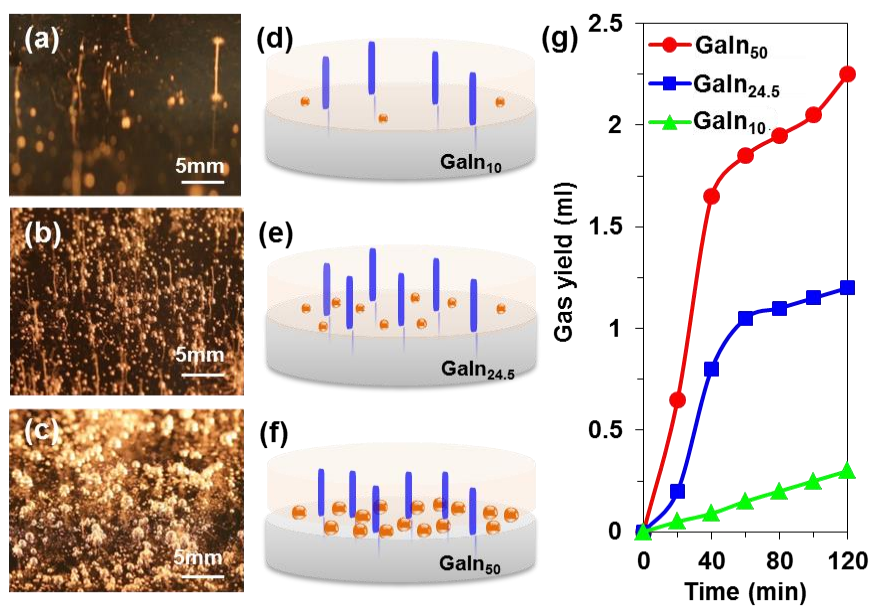

FIG.2. Gas plumes and gas bubbles generation at the interface of Ga-In alloy and NaOH electrolyte and gas yields for different In ratios. (a)-(c) Images of gas plumes form at the interface of GaIn$_{10}$, GaIn$_{24.5}$, and GaIn$_{50}$ in 1 M NaOH solution, respectively. (d)-(f) Schematic images of GaIn$_{10}$, GaIn$_{24.5}$, and GaIn$_{50}$ immersed in 1 M NaOH solution, respectively, illustrating the number of gas plumes and degree of bubbles formation. Every blue column represents a gas plume, and every orange dot represents a gas bubble. (g) Gas yield curves of the three Ga-In alloys over time.

In general, when Ga or In reacts with a strong base solution, the gas generation reaction may follow a well-defined chemistry reaction, taking Ga for example,[20] the reaction is shown as:

$$2Ga+2NaOH+2H_2O \rightarrow 2NaGaO_2+3H_2\uparrow \qquad (1)$$

However, the underlying mechanisms of gas plume and gas bubble formation, together with the different rates of gas generation in the various types of Ga-In alloy remains unclear. First of all, to exclude the possibility of impurities involved in the gas generation, the amount of Ga, In, together with Al, Cu, Fe, Ni, Pb and Zn in the electrolyte has been determined by ICP-OES after the gas forms for 72 h over GaIn$_{24.5}$. The dissolved amount of Ga and In (99.96%) are far greater than that of Al, Cu, Fe, Ni, Pb and Zn [0.04% in total, see Table.S1], which rules out the possibility of impurities-based redox couples for the Ga-In alloy. Next, we confirm that when Ga-In alloy is exposed to air before adding base solution, a dense passivation film (PF) [21] is rapidly formed on its surface due to its intrinsic chemical property [Fig.3(a)], which protects the Ga-In alloy from further oxidation (Multimedia view). However, this passivation layer also hinders the surface activity of the Ga-In alloy. The oxidation film was characterized by X-ray photoelectron spectroscopy (XPS) analysis,[22, 23] which indicates that the film is mainly composed of $Ga_2O_3$, $In_2O_3$, $Ga_2O$ (not stable phases) [Fig.3(b)&3(c)].

With alkaline solution transferring to liquid metal with PF, we assume that breaking of the passivation layer on the Ga-In alloy surface is critical to gas generation. To confirm this assumption, the gas eruption phenomenon of GaIn$_{24.5}$ was investigated in various pH environments. Typical results are shown in Fig.3(d) and a schematic illustration of the gas eruption in solutions with different pH was shown in Fig.S2. The dynamic process is shown in the Multimedia view. No gas plumes or bubbles occur in the pH range from 2 to 10. Weak acid or base is not strong enough to break the oxidation film and no gas generates. However, in strongly acidic or basic environments, the passivation layer could be disrupted and gas starts to blow out. And it is further found that the gas plume that appears in 1 M NaOH (pH= 14 ) is more dense than that appears in 1 M HCl (pH= 0 ) [Fig.3(d)]. Compared with the rapid venting of gas bubbles in the basic environment, acid conditions allowed separate small gas bubbles to be seen, the gas bubble is less vigorous in the acid environment in total.

The difference in reactivity can be attributed to the different rate of chemical reactions of the oxidation film with acidic and basic electrolyte. The amount of the exposure of liquid metal to electrolyte is closely related to the electrode potential of the reactants. The open circuit potential (OCP) of eutectic Ga-In alloy (GaIn$_{24.5}$) in DI water, 1 M NaOH and 1 M HCl solutions was investigated by electrochemical methods (experimental device is shown in Fig. S3, measurements of OCP are given in supplementary methods) with typical outputs presented in Fig. 3(e)&3(f). The absolute OCP of GaIn$_{24.5}$ in 1 M NaOH is higher than that in 1 M HCl. It has been proven that OCP $E^O$ has a positive correlation with chemical equilibrium constant $K$, basing on the Nernst Equation $\ln K = \frac{nFE^O}{RT}$ under equilibrium condition,[24] where $R$ is ideal gas constant, $F$ is Faraday's constant, $n$ is mole number, $T$ is temperature. Therefore, it indicates that 1 M NaOH can more effectively dissolve the oxidation film than 1 M HCl is able to. The increased exposure possibility of liquid metal alloy to alkaline electrolyte may induce the rise in the number of gas plumes and bubbles.

Compared with the steady OCP of GaIn$_{24.5}$ in strong base and acid solution, the OCP of GaIn$_{24.5}$ in deionized water undergoes a clear periodic change, which is attributed to the oxidation and removal of oxides of drop gallium electrode in deionized water, hence, confirmed the existence of the oxidation film [Fig.3(e)]. To further display the stability of the film in different electrolytes, I-V characteristics were also obtained by using an electrochemical station (method is given in supplementary information, experimental device is shown in Fig.S4). It is found that the I-V characteristics remained unchanged within the measurement range in DI water (0~-2V) [Fig. S5(a)], which also shows the stable existence of the oxide film. In the presence of 1 M HCl, I-V curves of GaIn$_{24.5}$ demonstrates a little increase compared to that in DI water, which may indicate slight damage in oxide film in 1 M HCl [Fig.S5(b)&Fig.S5(c)]. In contrast, there is a significant change and strong response in the form of the I-V curves in 1 M NaOH [Fig.S5(b)], which is expected due to damage of the film and the existence of electrochemically active substances in electrolyte. It is proved that destruction of PF in varying degree greatly affects gas generation rate.

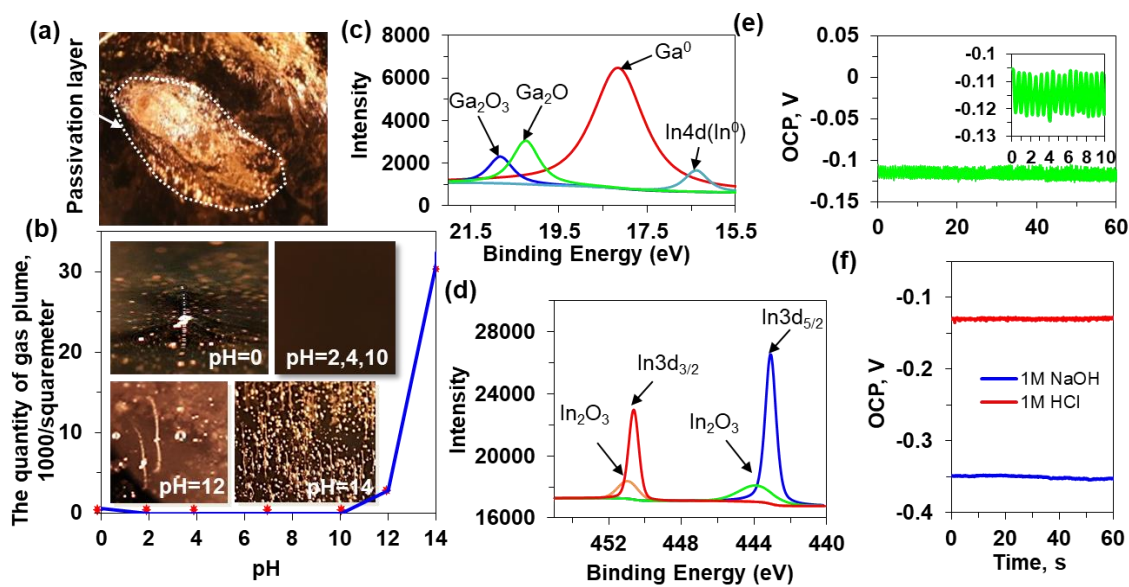

FIG.3. Different rates of reaction in solutions with various pH values showing the existence of a passivation film. (a) Formation of a passivation film on the surface of GaIn$_{24.5}$. (Multimedia view) XPS characterization of passivation film showing the existence of (b) Ga$_2$O$_3$ and (c) In$_2$O$_3$. (d) Change of the number of gas eruptions from GaIn$_{24.5}$ at different solution pH. (Multimedia view) (e) OCP of GaIn$_{24.5}$ in deionized water demonstrating the existence of an oxidation film. (f) OCP comparison of GaIn$_{24.5}$ in 1 M NaOH (pH = 14) and 1 M HCl (pH = 0).

Additionally, it is shown that the evolution of gas at the surface of the Ga-In alloy could continue for days. The gas generated in the reaction chamber was characterized by gas chromatography (GC) after the reaction of the Ga-In alloy and 1 M NaOH reaction for 12 h (Fig. S6) and 120 h [Fig.4(a)], respectively. We confirm that hydrogen is generated during the gas eruption [Fig.4(a)&Fig.S6], and that hydrogen gas production increased gradually with reaction time. The ratios of H$_2$/N$_2$ are 0(air), 0.06(GaIn$_{10}$), 0.27(GaIn$_{24.5}$), and 0.40(GaIn$_{50}$) at 12 h, and 0(air), 1.07(GaIn$_{10}$), 10.13(GaIn$_{24.5}$), and 18.42(GaIn$_{50}$) after 120 h, respectively. Detailed GC analysis further demonstrates that the peak-area ratio of $N_2/O_2$ after day 5 of the reaction are 3.32(air), 3.24(GaIn$_{10}$), 3.18(GaIn$_{24.5}$), and 3.12(GaIn$_{50}$), respectively. These results indicate that oxygen also generated between GaIn$_{24.5}$ and 1 M NaOH [Fig.4(a)].

The PF dissolves in the NaOH solution ($Ga_2O_3+2NaOH=2NaGaO_2+H_2O$),[25] which exposes the Ga-In alloy to the electrolyte. As electrode, Ga-In alloy is more active than PF, once alloy and PF are connected in the electrolyte solution, galvanic cell is formed. In this system, the exposed Ga-In alloy is electron rich and behaves as an anode,[26] where reduces hydrogen ions to hydrogen. When the distance between atoms is increased, the forbidden energy band narrows,[27] which abridges the distance from the conduction band to the valence band. Valence electrons become free electrons and transfer to the liquid metal, driven by the hydrogen generation occurring at concentrated stress point of the passivation film. The surface passivation film, lacking electrons, serves as the cathode which oxidizes water or hydroxyl ions to oxygen. Thus, once alloy and PF are connected in the electrolyte solution, galvanic cell is formed. The lighter hydrogen requires less buoyancy and is easy to float to the surface of electrolyte, then forms a gas plume. However, the oxygen bubbles, due to its poor wettability, usually stick [28] to the surface of PF when begin to form, and need to expand to a certain volume before floating to the surface [see Fig.S1]. This is consistent with the observed volcanic eruption phenomenon. Our results indicate electrolysis of water by a galvanic interaction [Fig.4(b)], as typically described by equation:[26]

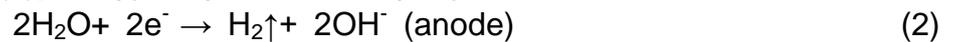
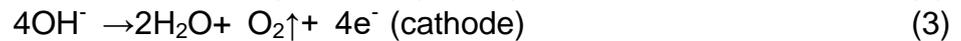

$$2H_2O + 2e^- \rightarrow H_2\uparrow + 2OH^- \quad \text{(anode)} \quad (2)$$
$$4OH^- \rightarrow 2H_2O + O_2\uparrow + 4e^- \quad \text{(cathode)} \quad (3)$$

According to Oshima et al.,[29] the volume of $O_2$ generated at semiconductor-liquid interfaces is generally substoichiometric which is also consistent with our results. This may be attributed to the fact that the potential of the hydrogen being lower than the standard reduction potential of Ga-In alloy and oxygen,[5] which makes oxidization of Ga or In and oxygen reduction reaction both prior to the evolution of hydrogen. It also explains the interesting phenomena observed in our experiment that large bubbles occurring next to gas plumes.

As the proportion of indium in the alloy is increased, a larger amount of gas developed. To explain this phenomenon, XRD analysis of the passivation film of GaIn$_{10}$, GaIn$_{24.5}$ and GaIn$_{50}$ has been conducted [Fig.4(c)]. All the three samples have been characterized as monoclinic structure. The pseudo-cubic lattice parameters of samples are 3.84 Å (GaIn$_{10}$), 3.93 Å (GaIn$_{24.5}$) and 3.97 Å (GaIn$_{50}$), respectively. The result is close to the computing result based on density functional theory and Vegard's law reported in Peelaers' work: 3.80 Å(GaIn$_{10}$), 3.85 Å(GaIn$_{24.5}$), 3.96 Å(GaIn$_{50}$).[30] As the proportion of In increased, the oxidation product $In_2O_3$ also increases [Fig.S7]. Larger lattice distortion happens due to the increase in the substitution of the $GaO_6$ by the bigger $InO_6$ octahedra [31] in the structure of gallium oxide.

We found that the band gap of semiconductor can be described as $E_g = \frac{\hbar^2}{48m_0 a^2}$, where $m_0$ refers to inertia mass of the cavities, and $\hbar$ is the reduced Planck constant. Thus, the band gap of the semiconductor ($E_g$) will be lowered by the expansion of lattice parameters (*a*) in the passivation film. Moreover, the band gap of Ga$_x$In$_{2-x}$O3 determined[31] and concluded[32] in previous work agrees with this trend: Eg[(Ga$_{1-x}$In$_x$)$_2$O$_3$]=(1-x)Eg[Ga$_2$O$_3$]+xEg[In$_2$O$_3$], where

Eg[$Ga_2O_3$]=4.8 eV and Eg[$In_2O_3$]=2.5 eV.

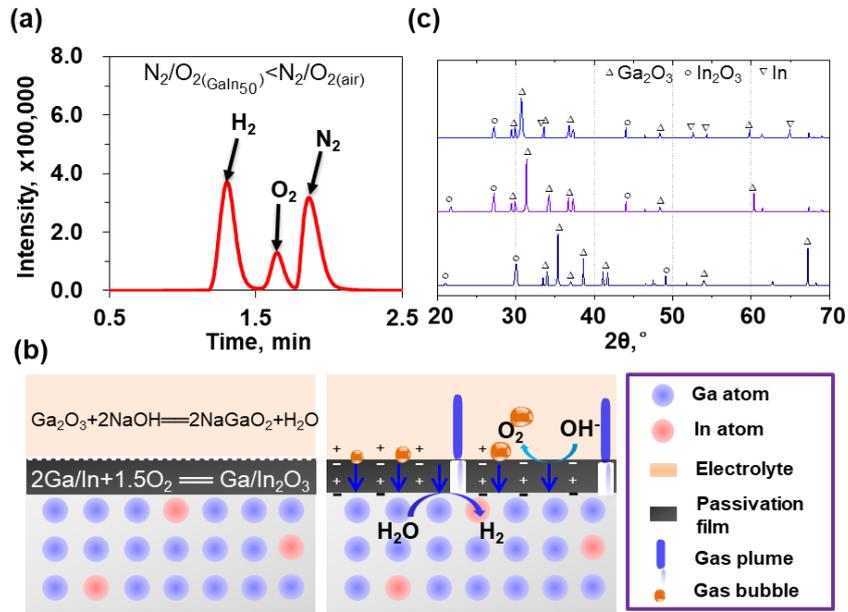

**FIG.4**. Gas chromatography analysis and schematic illustration of hydrogen and oxygen formation. (a) Typical gas chromatography characterization of gas after GaIn$_{50}$ and 1 M NaOH reacting for 120 h showing hydrogen and oxygen peaks. The peak-area ratio of $N_2/O_2$ after 5-day reaction for GaIn$_{50}$ is 3.12, i.e., smaller than that of air (3.32). (b) Schematic illustration of gas eruption mechanism. (c) XRD analysis of the passivation film of GaIn$_{10}$, GaIn$_{24.5}$ and GaIn$_{50}$.

When the energy level $E > E_g$, the density of carriers that can transfer in the passavation layer can be described as:

$$n = \int_{E_g}^{\infty} g(E)f(E)\,dE \qquad (4)$$

where, $g(E)$ is the capacity of the carrier state density at a certain energy level $E$, $f(E)$ is distribution probability of carriers at a certain energy level E. As $E_g$ decreases, the density of carriers $n$ rises.

The current density $J = en\bar{v}$ [32] increases as $n$ is increased, and the relationship between the oxygen generation rate($v$) and the density of cavity transfer can be defined by: $v_{O_2} = \gamma J$, where γ is a constant. The greater the oxygen generation rate, the faster the filling rate of the cavity in the oxide film. As more electrons are lost from the oxide film, more electrons transfer to the metal, which contribute to hydrogen generation. The rate of hydrogen generation can be described as: $v_{H_2} = \alpha v_{O_2}$, where α is a constant.

In conclusion, we discovered volcanic eruption-like phenomena which displayed from Ga-In liquid metal in electrolyte. The gas plume, occurring from the interior of liquid metal, resulted from the evolution of hydrogen and large bubbles, on the film surface, resulted from the generation of oxygen. In addition, the gas generation rate increases when the ratio of In is elevated. According to the above discussions, the gas eruption appears to be the result of the extraordinary properties of the passivation film, which dual contacting with the liquid metal and the electrolyte. The system can be regarded as a combination of a galvanic interaction and a semiconductor band structure, where a huge band barrier separates the valence and conduction bands and a steady flow of electrons connects the semiconductor and liquid metal. This finding is expected to be important for developing room temperature liquid metal machine and hydrogen generation technology.

**Supplementary Material**

See supplementary material for the supplementary methods, supplementary multimedia video (S1& S2), supplementary figures (S1-S7) and supplementary table (S1).

**Acknowledgements**

We acknowledge Professor Wenfu Fu and Dr. Fang Li from Technical Institute of Physics and Chemistry, CAS for their help with gas chromatography and I-V curve detection. We acknowledge Lzmarie Poventud-Fuentes from the University of Pennsylvania for help editing the manuscript. This research is supported by the Dean's Research Funding (No. 2015-LJ), the Frontier Project from Chinese Academy of Sciences (No. QYZDJ-SSW-JSC016) and TIPC Director Fund (No. Y6AL021R2X).

**Author Contributions**

W. Rao and J. Liu conceived the project. R．Zhao conducted and designed gas eruption experiments, H. Wang conducted XPS experiments, J. Tang helped with W. Rao on the measurement of open circuit potential and I-V curve, W. Rao also performed gas yield and chromatography characterization. R. Zhao, W. Rao, and J. Liu wrote the manuscript with contributions from all authors, and all authors reviewed and approved the manuscript.